\documentclass[manuscript]{aastex}

\usepackage{graphicx,bm,color}
\usepackage{amsmath}
\usepackage{amssymb}
\usepackage{amsfonts}

\newcommand{\be}{\begin{equation}}
\newcommand{\ee}{\end{equation}}
\newcommand{\bea}{\begin{eqnarray}}
\newcommand{\eea}{\end{eqnarray}}
\newcommand{\beaa}{\begin{eqnarray*}}
\newcommand{\eeaa}{\end{eqnarray*}}

\newcommand{\nn}{\nonumber \\}
\newcommand{\e}{\mathrm{e}}

\allowdisplaybreaks[4]

\shorttitle{Inflation without self-reproduction in $F(R)$ gravity}
\shortauthors{Nojiri and Odintsov}

\begin{document}

\title{Inflation without self-reproduction in $F(R)$ gravity}

\author{Shin'ichi Nojiri}
\affil{
Kobayashi-Maskawa Institute for the Origin of Particles and the
Universe, Nagoya University, Nagoya 464-8602, Japan \\
and \\
Department of Physics, Nagoya University, Nagoya 464-8602, Japan \\
}

\and

\author{Sergei D. Odintsov}
\affil{
Consejo Superior de Investigaciones Cient\'{\i}ficas, ICE/CSIC-IEEC,
Campus UAB, Facultat de Ci\`{e}ncies, Torre C5-Parell-2a pl, E-08193
Bellaterra (Barcelona), Spain \\
and \\
Instituci\'{o} Catalana de Recerca i Estudis Avan\c{c}ats
(ICREA), Barcelona, Spain
}

\begin{abstract}

We investigate inflation in frames of two classes of $F(R)$ gravity and 
check its consistency with Planck data. It is shown that $F(R)$ inflation
without self-reproduction may be constructed in close analogy with the 
corresponding scalar example proposed by Mukhanov for the resolution the 
problems of multiverse, predictability and initial conditions.

\end{abstract}



\maketitle

\section{Introduction}

Recent observational results from different experiments indicate
towards the very natural choice of inflation as the very-early universe epoch.
Still, specific inflationary models are not free from various problems.
One problem was indicated in Ref.~\citep{Mukhanov:2014uwa} in relation with
cosmological perturbation theory (for the introduction and review, see 
\citep{Mukhanov:2005sc,Mukhanov:2013tua,R}).
In order to resolve the problem of multiverse, predictability and
initial conditions it has been proposed the inflationary model without
self-reproduction. The explicit scalar inflationary theory without
self-reproduction has been formulated there \citep{Mukhanov:2014uwa}. 
Note that different approach to the resolution of above problems of 
multiverse and initial conditions was also developed (for a review, 
see, \citep{Linde:2014nna}).

In order to avoid the self-reproduction of the universes at the inflationary
epoch,
Mukhanov proposed the following conditions \citep{Mukhanov:2014uwa}:
\begin{enumerate}
\item\label{a} $1+w (N) = 1$ at $N = 1$ (to have graceful exit),
\item\label{b} $1+w (N) \leq 2/3$ at $N = N_m$ (to solve initial condition
problem),
\item\label{c} $1+w (N) \ll 1$ for $1 < N < N_m$ (inflation),
\item\label{d} $1+w (N) > \epsilon (N)$ for $1 < N < N_m$ (no
self-reproduction).
\end{enumerate}
Here $N$ is the e-folding defined by $a=a_0 \e^{-N}$ with a constant $a_0$
and $w(N)$ is the equation of state parameter.
Furthermore, $\epsilon(N)$ is the energy-density of the Planck unit and it is
related with
the energy-density of the usual unit by
$\epsilon(N) \sim \kappa^4 \rho(N)$.
The purpose of this letter is
to construct an inflationary model of $F(R)$ gravity satisfying the condition
of no self-reproduction. 
See \citep{Capozziello:2014hia} for the cosmology from the inflation to the dark energy 
in the $F(R)$ gravity. 
Note that we consider $F(R)$ frame, so our discussion
is not applied to modified gravities which have no inflationary epoch as direct
solution of dynamical equations.

\section{The explicit inflationary model of $F(R)$ gravity.}
First, we should note that if there is no matter, we may consider the effective
EoS
parameter by
\be
\label{M1}
w = - 1 - \frac{2 \dot H}{3 H^2} = - 1 - \frac{2 H'}{3H}\, .
\ee
Here $H$ is the Hubble rate defined by $H=\dot a/a$.
Therefore
\be
\label{M2}
H = H_0 \e^{ - \frac{3}{2} \int^N dN' \left( 1 + w \left( N \right) \right)}\,
.
\ee
Especially when $1 + w (N) = 1$, $H \propto \e^{ - \frac{3}{2} N}$ and
when $1 + w(N) = 2/3$, $H \propto \e^{ - N}$.

It is well-known that when $F(R) \propto R^n$,
\be
\label{M3}
H \propto \e^{ - N/\alpha}\, , \quad
\alpha = \frac{ - 1 + 3n - 2n^2}{n-2}\, .
\ee
Then when one regards $w$ as a constant, we find
\be
\label{M4}
n = \frac{ - \alpha + 3 \pm \sqrt{\alpha^2 + 10 \alpha + 1}}{4}\, .
\ee
Note that if $\alpha>0$, $n$ is real.
Especially when $1 + w (N) = 1$, that is, $\alpha=2/3$,
\be
\label{M5}
n = n^f_\pm \equiv \frac{ 7 \pm \sqrt{73}}{12} = 1.29533...\,, \ -
0.12866...\, ,
\ee
and when $1 + w(N) = 2/3$, that is, $\alpha=1$,
\be
\label{M6}
n = \frac{1 \pm \sqrt{3}}{2} = 1.366026...\,, \ -0.36602...\, .
\ee
Let define
\be
\label{M7}
n^i_\pm \equiv \frac{ - \alpha + 3 \pm \sqrt{\alpha^2 + 10 \alpha + 1}}{4}\, ,
\ee
when $1 + w(N) \leq 2/3$, that is, $\alpha \geq 1$.
Note that $n_+^i$ and $n_+^f$ are positive but
$n_-^i$ and $n_-^f$ are negative.

First, we may consider the following model
\be
\label{M8}
F(R) = F_i \left( \frac{R}{R_e} \right)^{n^i_+} + F_f \left( \frac{R}{R_e}
\right)^{n^f_-}\, .
\ee
Here we do not include the Einstein-Hilbert term $R/2\kappa^2$. The above model
belongs to the class of theories with positive and negative powers of
curvature. Such theories have been introduced in Ref.~\citep{NO2003} in order to
unify inflation with dark energy (for general review of the unification of 
inflation with dark energy in $F(R)$ cosmology, see \citep{Nojiri:2006ri,Nojiri:2010wj,Capozziello:2011et,Capozziello:2010zz}).
When $R\gg R_e$, the first term dominates because $n^i_+$ is positive and
the condition \ref{b} could be satisfied.
On the other hand, when $R\ll R_e$, the second term dominates because $n^i_+$
is negative
and the condition \ref{a} could be satisfied.
In order to avoid the anti-gravity, we should require $F'(R)>0$, which shows
that $F_i>0$ and $F_f<0$.

If there is a solution $R=R_0$ for the equation $\frac{d}{dR}
\left(\frac{F(R)}{R^2}\right)=0$,
$R=R_0$ corresponds to the (anti-)de Sitter space-time.
In case of the model (\ref{M8}), one gets \be
\label{M9}
R_0 = R_e \left\{ - \frac{ \left( n^i_+ - 2 \right) F_i }{ \left( n^f_- - 2
\right) F_f }
\right\}^{ - \frac{1}{n^i_+ - n^f_- }}\, .
\ee
Note that
\be
\label{M10}
  - \frac{ \left( n^i_+ - 2 \right) F_i }{ \left( n^f_- - 2 \right) F_f } > 0
\, .
\ee
In the de Sitter space-time, if $R=12 H_0^2$, we have $H=H_0$.
Around the de Sitter solution, $F(R)$ in (\ref{M8}) can be approximated by
\be
\label{M11}
F(R) \sim R^2 \left( F_0 + F_1 \left( R - R_0 \right)^2 \right)\, .
\ee
Here
\begin{align}
\label{M12}
F_0 \equiv & \frac{n_+^i - n_-^f}{2 - n_-^f}
F_i \left\{ - \frac{ \left( n^i_+ - 2 \right) F_i }{ \left( n^f_- - 2 \right)
F_f }
\right\}^{ - \frac{n^i_+}{n^i_+ - n^f_- }}\, , \\
\label{M13}
F_1 \equiv & \left(n_+^i - 2\right)\left( n_+^i - n_-^f\right)
F_i \left\{ - \frac{ \left( n^i_+ - 2 \right) F_i }{ \left( n^f_- - 2 \right)
F_f }
\right\}^{ - \frac{n^i_+}{n^i_+ - n^f_- }} \frac{1}{R_0^2} \nn
=& - \left(n_+^i - 2\right)\left( n^f_- - 2 \right) \frac{F_0}{R_0^2}
\, .
\end{align}
One may consider the pertubation around the de Sitter space-time as $H=H_0 +
H_1 \left(N\right)$.
Then by using the following equation, which corresponds to the first FRW
equation,
\be
\label{M13B}
0 = - \frac{1}{2} F(R) + 3 \left( H^2 + H H' \right) F'(R)
  - 18 \left( 4 H^4 H' + H^2 \left( H' \right)^2 + H^3 H'' \right) F'' (R)\, ,
\ee
we obtain the equation linearized with respect to $H_1(N)$,
\be
\label{M14}
0 = 4 R_0^2 F_1 H_1(N) + 3 \left( F_0 - R_0^2 F_1 \right) H_1'(N)
  - \left( F_0 + R_0^2 F_1 \right) H_1''(N)\, ,
\ee
whose solution is given by
\be
\label{M15}
H_1 (N) = C_+ \e^{\lambda_+ N} + C_- \e^{\lambda_- N} \, .
\ee
Here $C_\pm$ are constants and
\be
\label{M16}
\lambda_\pm = \frac{ - 3 \left( F_0 - R_0^2 F_1 \right)
\pm \sqrt{ 9 F_0^2 - 2 R_0^2 F_0 F_1 + 25 R_0^4 F_1^2}}{ -2 \left ( F_0 + R_0^2
F_1
\right)}\, .
\ee
If we assume $\left| R_0^2 F_1 \right| \ll \left| F_0 \right| $, we find
\be
\label{M17}
\lambda_+ \sim 3\, , \quad \lambda_- \sim - \frac{4}{3} R_0^2 \frac{F_1}{F_0}\,
.
\ee
The $\lambda_+$ corresponds to the rapidly growing mode and $\lambda_-$
to the slowly growing mode.
If one can choose $C_+=0$ by a boundary condition, the period of the inflation
in terms of the e-folding $N$ is given by
\be
\label{M18}
N_m 
\sim \frac{1}{ \left| \lambda_- \right|}
\sim \left| - \frac{3 F_0}{4 R_0^2 F_1}\right|
= \left| \frac{3}{4} \left(n_+^i - 2\right)\left( n^f_- - 2 \right)\right| \,
.
\ee
We should require a condition $N_m \sim 60-70$ but Eq.~(\ref{M18}) shows that
the condition
cannot be satisfied and therefore the condition \ref{c} also cannot be
satisfied.

About the condition (\ref{d}), we may also regard $\epsilon \sim \kappa^2 R_e$.
Eq.~(\ref{M15}) with $C_+=0$ indicates \be
\label{M19}
1 + w(N) \sim \frac{H'}{H} \sim \frac{C_- \lambda_- \e^{\lambda_- N}}{H_0}\, .
\ee
Therefore if we choose
\be
\label{M20}
\frac{C_- \lambda_- }{H_0} \gg \kappa^2 R_e\, ,
\ee
the condition (\ref{d}) could be satisfied.

Since the model (\ref{M8}) does not satisfy the condition (\ref{c}), we modify
the model as
follows,
\be
\label{M8B}
F(R) = \frac{1}{R^2} \left( F_i \left( \frac{R}{R_e}
\right)^{\frac{n^i_+-2}{\alpha}}
+ F_f \left( \frac{R}{R_e} \right)^{\frac{n^f_- - 2}{\alpha}} \right)^\alpha
\, .
\ee
Here $\alpha$ is a constant. 
The behavior when $R\gg R_e$ and $R\ll R_e$ does not change from those in the
model (\ref{M8}) and therefore
the conditions (\ref{a}) and (\ref{b}) can be satisfied again.
The condition (\ref{d}) is also satisfied with the choice (\ref{M20}).
Therefore we only need to check the condition (\ref{c}).
Instead of (\ref{M9}), the solution describing the de Sitter space-time is
given by
\be
\label{M9B}
R_0 = R_e \left\{ \left( - \frac{n^i_+ - 2}{n^f_- - 2} \right)^\alpha
\frac{F_i }{F_f }\right\}^{ - \frac{1}{n^i_+ - n^f_- }}\, .
\ee
Around the de Sitter solution, $F(R)$ in (\ref{M8B}) can be approximated
as in (\ref{M11}) but instead of (\ref{M12}), we find
\begin{align}
\label{M12B}
F_0 =& \left( n_+^i - n_-^f \right)^\alpha \left( \frac{F_i}{ \left( - n^f_- + 2
\right)^\alpha}
\right)^{\frac{n^f_- - 2}{n^f_- - n^i_+}}
\left( \frac{F_f}{ \left( n^i_+ - 2 \right)^\alpha}
\right)^{\frac{n^i_+ - 2}{n^i_+ - n^f_-}}\, , \nn
F_1 =& \frac{n^i_+ + n^f_- - 4 - 4 \alpha}{2\alpha R_0^2}\frac{F_0}{R_0^2}\, .
\end{align}
Then instead of (\ref{M18}), one obtains
\be
\label{M18B}
N_m 
\sim \frac{1}{\left| \lambda_-\right|}
\sim \left| - \frac{3 F_0}{4 R_0^2 F_1} \right|
= \left| - \frac{3 \left( n^i_+ + n^f_- - 4 - 4 \alpha \right)}{8\alpha}
\right| \, .
\ee
By adjusting $\alpha$, we obtain $N_m \sim 60-70$ and the condition (\ref{c})
can be satisfied.

Let us estimate the scalar index $n_\mathrm{s}$ of the curvature perturbations
and the tensor-to-scalar ratio
$r$ of the density perturbation, and the running of the spectral index
$\alpha_\mathrm{s}$.
The corresponding expressions for $F(R)$ gravity are given in 
\citep{Bamba:2014wda}.
Using the expression of $H$ in (\ref{M15}) with $C_+=0$, we find the following
expressions for the slow-roll
parameters, $\epsilon$, $\eta$, $\xi^2$ \citep{Bamba:2014daa},
\be
\label{eex}
\epsilon= - \lambda_-\, , \quad \eta = - 2 \lambda_-\, , \quad \xi^2 = 4
\lambda_-^2\, .
\ee
Therefore,
\begin{align}
\label{eex2}
n_\mathrm{s} - 1 =& - 6 \epsilon + 2\eta = 2\lambda_- +
\mathcal{O}\left(\lambda_-^2 \right)\, , \quad
r = 16 \epsilon = - 16 \lambda_- + \mathcal{O}\left(\lambda_-^2 \right)\, ,
\nn
\alpha_\mathrm{s} =& 16 \epsilon \eta - 24 \epsilon^2 - 2 \xi^2 =
\mathcal{O}\left(\lambda_-^3 \right)\, .
\end{align}
The Planck data~\citep{Ade:2013lta,Ade:2013uln} suggest
$n_{\mathrm{s}} = 0.9603 \pm 0.0073\, (68\%\,\mathrm{CL})$,
$r< 0.11\, (95\%\,\mathrm{CL})$,
and $\alpha_\mathrm{s} = -0.0134 \pm 0.0090\, (68\%\,\mathrm{CL})$
[the Planck and WMAP~\citep{Spergel:2003cb,Spergel:2006hy,Komatsu:2008hk,Komatsu:2010fb, Hinshaw:2012aka}], the negative sign of
which is at $1.5 \sigma$.
The data $n_\mathrm{s} \sim 0.9603$ show that $1/\lambda \sim - 50$ but
$r\sim 0.11$ indicates that $1/\lambda \sim - 145$.
This discrepancy always occurs when we consider the linearized model in
(\ref{M14}) because this discrepancy is
due to the exponential behavior in the solution (\ref{M15}).
Therefore, if we include non-linear corrections, the above values could be
improved.

We may rewrite the action (\ref{M8}) or (\ref{M8B}) in a scalar-tensor form by
introducing a new scalar field $\sigma$ by
\be
\label{M21}
\sigma = - \ln \left( 2\kappa^2 F'(R) \right)\, .
\ee
Here, it is introduced $2\kappa^2$.
We now consider a case $R\gg R_e$ and another case $R\ll R_e$.
When $R\gg R_e$, one finds
\be
\label{M23}
\sigma \sim - \left( n^i_+ - 1 \right) \ln \frac{R}{R_e} - \ln \left(2 n^i_+
F_i \kappa^2 \right)\, ,
\ee
and the corresponding scalar-tensor theory looks as 
 (for general review of scalar-tensor gravity, see 
\citep{Fujii:2003pa,Faraoni:2004pi})
\begin{align}
\label{M24}
S =& \frac{1}{2\kappa^2} \int d^4 x \sqrt{-g} \left( R - \frac{3}{2}
\partial_\mu \sigma
\partial^\mu \sigma - V(\sigma) \right)\, , \nn
V(\sigma) \sim& \frac{R\left(\sigma\right)}{F'\left( R \left( \sigma \right)
\right)}
  - \frac{F \left( R\left(\sigma\right) \right)}{F'\left( R \left( \sigma
\right) \right)^2}
= \frac{n^i_+ - 1}{n^i_+} \frac{2\kappa^2 R_e \e^{\frac{n^i_+ - 2}{n^i_+ - 1}
\sigma}}{
\left( 2 n^i_+ F_i \kappa^2 \right)^{\frac{1}{n^i_+ - 1}}}\, .
\end{align}
On the other hand, when $R\ll R_e$, we find
\be
\label{M25}
\sigma = - \left( n^f_- - 1 \right) \ln \frac{R}{R_e} - \ln \left(2 n^f_- F_f
\kappa^2 \right)\, ,
\ee
and
\be
\label{M26}
V(\sigma)
= \frac{n^f_- - 1}{n^f_-} \frac{2\kappa^2 R_e \e^{\frac{n^f_- - 2}{n^f_- - 1}
\sigma}}{
\left( 2 n^f_- F_f \kappa^2 \right)^{\frac{1}{n^f_- - 1}}}\, .
\ee
Furthermore, we also consider the case that $R\sim R_0$ and $F(R)$ in
(\ref{M8}) or
(\ref{M8B}) can be approximated by (\ref{M11}).

In  $F(R)$ gravity, both of the tensor mode and the scalar mode in the
metric $g_{\mu\nu}$ appear as
propagating modes.
Rewriting $F(R)$ gravity in the scalar-tensor form, we can separate the tensor
mode and the scalar mode
and find the Newton law in  $F(R)$ gravity by the tensor mode is identical
with that in the Einstein gravity
although the coupling depends on the value of the scalar in the background.
The propagation of the scalar mode gives an additional correction to the Newton
law.

The mass of $\sigma$ is \be
\label{JGRG24}
m_\sigma^2 \equiv \frac{3}{2}\frac{d^2 V(\sigma)}{d\sigma^2}
=\frac{3}{2}\left\{\frac{R}{F'(R)} - \frac{4F(R)}{\left(F'(R)\right)^2} +
\frac{1}{F''(R)}\right\}\, .
\ee
Then if $m_\sigma$ is not large enough, the large correction to the Newton law
appears in general.
For the model (\ref{M8}) or (\ref{M8B}), when $R\ll R_e$, we find
\be
\label{AAA1}
F(R) \sim F_f \left( \frac{R}{R_e} \right)^{n^f_-}\, .
\ee
Hence, $m_\sigma^2 \sim R$.
In the present universe, the order of the mass $m_\sigma$ should be that of the
Hubble rate, $m_\sigma \sim H \sim 10^{-33}\,\mathrm{eV}$,
which is very light and could make the correction to the Newton law very large.

In \citep{Hu:2007nk}, realistic $F(R)$ model was proposed.
It has been found, however, that the model has an instability where
the large curvature can be easily produced (manifestation of a
possible future singularity).
In the model of \citep{Hu:2007nk}, a parameter
$m\sim 10^{-33}\, \mathrm{eV}$ with a mass dimension is included.
The parameter $m$ plays a role of the effective cosmological constant.
When the curvature $R$ is large enough compared with $m^2$, $R\gg m^2$,
$F(R)$ \citep{Hu:2007nk} looks as follows:
\be
\label{HS1}
F(R) = R - c_1 m^2 + \frac{c_2 m^{2n+2}}{R^n}
+ \mathcal{O}\left(R^{-2n}\right)\, .
\ee
Here $c_1$, $c_2$, and $n$ are positive dimensionless constants.
Similar viable models have been proposed in 
\citep{Appleby:2007vb,Nojiri:2008fk,Cognola:2007zu}.
Then it is possible to construct a model which behaves as (\ref{M8B}) at the
early universe but behaves as
(\ref{HS1}) at the present universe.
For this purpose, we define the following function of the scalar curvature $R$,
\be
\label{M27}
S_\pm \equiv \frac{1}{2} \left\{ 1 \pm \tanh \left( \frac{R}{R_m} -
\frac{R_m}{R} \right) \right\}\, .
\ee
Here $R_m$ is a constant which is much larger than the scalar curvature at the
present universe or the curvature
on the earth.
Let also $R_m$ is much smaller than the scale of the inflation, $R_m\ll R_e,\
R_0$.
Then we find that when $R \gg R_m$, $S_+ \to 1$, $S_-\to 0$, very rapidly and
when $R \ll R_m$, $S_+ \to 0$,
$S_-\to 1$.
One may consider the following model by using (\ref{M8B}):
\be
\label{M28}
F(R) = \frac{R}{2\kappa^2} S_- \left(R\right) + \left\{
\frac{1}{R^2} \left( F_i \left( \frac{R}{R_e} \right)^{\frac{n^i_+-2}{\alpha}}
+ F_f \left( \frac{R}{R_e} \right)^{\frac{n^f_- - 2}{\alpha}}
\right)^\alpha\right\} S_+(R)\, .
\ee
Then when $R\gg R_e$, the model (\ref{M8B}) can be reproduced.
On the other hand, when $R\ll R_e$, the Einstein gravity can be reproduced,
$F(R)\to \frac{R}{2\kappa^2}$.
Note that when $R\ll R_e$, which corresponds to the present universe, the mass
  (\ref{JGRG24}) for
the scalar mode is given by
\be
\label{M29}
m_\sigma^2 \sim \frac{3}{2} \frac{R^2}{R_m} \cosh^2 \left( - \frac{R_m}{R}
\right)\, ,
\ee
which is very large at the present universe or on the earth and the correction
to the Newton law becomes very
small.
Although the model does not generate the accelerating expansion at the present
universe, we can obtain the model
generating the accelerating expansion by replacing $\frac{R}{2\kappa^2}$ in the
first term in (\ref{M28}) with the
$F(R)$ in (\ref{HS1}).
Thus, we proposed $F(R)$ model which describes inflationary universe without 
self-reproduction and which behaves as General Relativity at weak curvature.

\section{Discussion.}

We may consider the condition (\ref{d}) for the self-reproduction of general
$F(R)$ gravity.
In $F(R)$ gravity, by using the formulation of the reconstruction
\citep{Nojiri:2006gh,Nojiri:2009kx}, one can construct the model reproducing the
following Hubble rate
\be
\label{M30}
H^2 = C f(N)\, .
\ee
Here $f(N)$ is an adequate function of the e-foldings $N$ and $C$ is a
constant.
Then the condition (\ref{d}) can be written as,
\be
\label{M31}
1+w (N) > \kappa^2 C f(N)\ \mbox{for}\ 1 < N < N_m \, .
\ee
Therefore if we choose $C$ to be small enough, the condition (\ref{d}) can be
always satisfied.
As an example, we consider the following model from Ref.~\citep{Bamba:2014wda},
\begin{align}
F(R)= & C_1(6G_0-2 R)^{3/2} \sqrt{\frac{R}{12 G_0}-\frac{1}{4}}
\left[1-\frac{1}{\frac{R}{12G_0}-\frac{1}{4}}-\frac{1}{4 \left(\frac{R}{12
G_0}-\frac{1}{4}\right)^2}\right] \nn
& +C_2 (6G_0-2 R)^{3/2} \text{L}\left(\frac{1}{2},\frac{3}{2};
\frac{R}{12G_0}-\frac{1}{4}\right)\, ,
\label{FRNe6}
\end{align}
which reproduces the Hubble rate,
\be
\left(H (N)\right)^2=G_0 N + G_1\, .
\label{eq:IIC1}
\ee
Here $G_0 (<0)$ and $G_1 (>0)$ are constants.
In (\ref{FRNe6}), $C_1$ and $C_2$ are constants of integration, $L(u_1,u_2;y)$
is the generalized Laguerre polynomial,
where $u_1$ and $u_2$ are constants and $y$ is a variable.
If we choose $(N, \kappa^2 G_0, \kappa^2 G_1) = (50.0, -0.850, 95.0)$ and
$(60.0, -0.950, 115)$, we obtain
$(n_\mathrm{s}, r, \alpha_\mathrm{s}) = (0.967, 0.121, -5.42 \times 10^{-5})$
and $(0.967, 0.123, -5.55 \times 10^{-5})$, respectively, which could be
consistent with the Planck data
although the model (\ref{eq:IIC1}) does not satisfy the condition (\ref{a})
because there is no the exit from the
inflation. Eventually, the exit should be described by another scenario, not
gravitational one or by adding of extra gravitational terms.
Comparing (\ref{M31}) and (\ref{eq:IIC1}), we find that if we choose $G_0$ and
$G_1$ small enough, the
condition (\ref{d}) is satisfied and the self-reproduction is prohibited.

In summary, we discussed two classes of $F(R)$ gravity which admits inflation 
in
$F(R)$ description. On the same time, such theory may be consistent with Planck 
data. This shows that gravitationally-induced $F(R)$ inflation
which avoids self-reproduction and resolves the problems of multiverse,
predictability and initial
conditions in the same sense as for scalar inflation is quite possible.

\acknowledgments

We are grateful to A.~Linde and V.~Mukhanov for very useful discussion.
The work is supported by the JSPS Grant-in-Aid for Scientific
Research (S) \# 22224003 and (C) \# 23540296 (S.N.) and in part by
MINECO(SPAIN), projects FIS2010-15640 and FIS2013-44881(S.D.O.).

\end{document}